\documentclass[11pt]{article}
\usepackage{graphicx}
\usepackage{graphics}
\usepackage{amssymb,amsmath,amsthm}
\usepackage{indentfirst}
\usepackage{color}
\usepackage[table]{xcolor}
\usepackage{hyperref}
\usepackage{enumerate}
\usepackage{fullpage}
\usepackage{authblk}
\usepackage{xr}

\def\bs{\boldsymbol}

\def\bsp{\begin{split}}
\def\esp{\end{split}}
\def\bit{\begin{itemize}}
\def\eit{\end{itemize}}
\def\benu{\begin{enumerate}}
\def\eenu{\end{enumerate}}
\def\bmat{\begin{bmatrix}}
\def\emat{\end{bmatrix}}

\date{}
\title{Dynamics of mosquito swarms over a moving marker}

\author[1,2]{Puneet Jain}
\author[3]{Om Prakash Singh}
\author[4]{Sachit Butail\thanks{Author for correspondence (sbutail@niu.edu)}}

\affil[1]{Department of Computer Science, Brigham Young University, UT, USA}
\affil[2]{Indraprastha Institute of Information Technology Delhi (IIIT-Delhi), New Delhi, India}
\affil[3]{National Institute of Malaria Research, Dwarka, New Delhi, India}
\affil[4]{Department of Mechanical Engineering, Northern Illinois University, IL, USA}

\begin{document}
\maketitle

\begin{abstract}
Insect swarms are a model system for understanding collective behavior where the collective motion appears in disorder. To initiate and maintain a swarm in place, flying insects often use a visual external cue called a marker. 
In mosquitoes, understanding the swarming behavior and its relation to the marker has an additional medical relevance since swarming often precedes mating in the wild, thus constituting an important stage to intercept for controlling mosquito population. In this paper, we conduct preliminary experiments to characterize the visual coupling between a swarm of mosquitoes and a marker. 
A laboratory microcosm with artificial lighting was built to stimulate consistent swarming in the malarial mosquito \emph{Anopheles stephensi}.
The experimental setup was used to film a mosquito swarm with a stereo camera system as a marker was moved back-and-forth with different frequencies. 
System identification analysis of the frequency response shows that the relationship between the swarm and the marker can be described by delayed second order dynamics in a feedback loop. Further, the length of the internal time delay appears to correlate with the number of mosquitoes swarming on the marker indicating that such a delay may be able  capture social interactions within swarming systems. For insect swarms, model fitting of trajectory data provides a way to numerically compare swarming behaviors of different species with respect to marker characteristics. These preliminary results motivate investigating linear dynamic system in feedback as a framework for modeling insect swarms and set the stage for future studies.

\end{abstract}

\maketitle


\section{\label{sec:level1}Introduction}
Among the variety of collective behaviors demonstrated by different animal species, swarming in flying insects occupies a unique position \cite{Dublon2014}. Unlike fish schools and bird flocks, which often appear to be coordinated, insect swarms appear disorganized. Yet, insect swarming is far from a random process as was shown in one of the first works by Okubo \cite{Okubo1986}, who studied collective motion of midge swarms and showed that swarming in insects is distinct from gaseous diffusion. With the ability to reconstruct three-dimensional motion, varying levels of coordination have been reported in mosquito swarms filmed in the wild \cite{Butail2013, Shishika2014}, and midge swarms filmed in the laboratory \cite{Ni2015, Puckett2015}. In terms of the order parameter that is often used to describe coordination in collective behaviors, insect swarms have been shown to lie at the edge of the phase transition between order and disorder \cite{Attanasi2014}.   

Among insects that swarm, the mosquito holds a special position in context of its role as a vector for many deadly diseases including malaria and dengue fever \cite{Benelli2016}. Mosquitoes often swarm during twilight over high contrast regions called markers \cite{Gibson1985, Downes1969, Sullivan1981}. These swarms are composed almost entirely of males with females entering at different times to mate \cite{Manoukis2009, Charlwood1980}. Females typically get inseminated only once \cite{Jones1973}, even though they lay eggs multiple times during their lifetime. From a malaria control perspective, this makes mosquito swarming, and the environmental cues that induce the same, an important feature to understand in order to intercept reproduction.  

Example of markers that induce swarms in the wild include a pile of trash or a patch of grass \cite{Gibson1985, Downes1969, Manoukis2009, Butail2013, Diabate2011}. Left undisturbed, these markers serve as consistent swarming sites every day, year after year, during the wet season \cite{Manoukis2009, Butail2013}. While experiments have been performed in the past to isolate the possible role of marker color, size, and movement on mosquito swarming \cite{Mohan1952, Nielsen1962}, most of these studies were purely observational; the visual marker was found to play a crucial role in eliciting different sized and shaped swarms in different species \cite{Poda2019}. More recently, van der Vaart et al. \cite{vanderVaart2019} use experiments with oscillating visual markers to characterize laboratory midge swarms in a rheological sense, showing the existence of high viscosity and low structure within such swarms. These properties were found to vary with frequency and contribute to the general damping of the external perturbations.  

From the perspective of collective behavior, the marker represents an environmental cue against which an individual in the swarm must weigh social influence. Since the swarm tends to follow the marker, a natural framework to consider modeling this behavior is in the form of a feedback loop that has been used to model several biological systems \cite{Cowan2014}. Within this framework, a dynamical system (swarm) responds to the difference between a reference input (marker movement) and its state (position relative to the marker). Furthermore, in our context, since we model the swarm as a single entity, we hypothesize that the effect of intra-swarm interactions is that it slows down the response of the swarm as a whole. This could be captured in the form of a delay, much in the same manner as the neural processing delay modeled in individual systems \cite{Sponberg2015}. A natural question to then consider is if this delay term is dependent on the swarm density. 

We address this question through the analysis of experiments where we filmed a mosquito swarm as it moved over a marker. This setup follows a classical approach in control systems engineering for identifying the properties of a mechanical or electrical system \cite{Ogata1997}. The approach involves perturbing a system (considered as a black box) with known periodic disturbances and observing the response. Accordingly, in our controlled laboratory experiments the marker was moved in a periodic motion (known disturbance) with different frequencies as we reconstructed the motion of the swarm (system) in three dimensions.

We hypothesized the following: (h1) the open-loop swarm dynamics within the feedback loop that best fit the data will be in the form of a second order system; this was based on prior work \cite{Butail2013} which showed that the interaction of mosquitoes within a swarm can be described as a damped harmonic oscillator;  (h2)  we further hypothesized that the best-fit swarm dynamics will consist of a non-zero delay term; and finally (h3) when the delay was allowed to vary between trials, we expected to see a dependence of the delay on the average number of positions tracked---in other words, more the number of mosquitoes following the marker, higher the intensity of social interaction and therefore longer the delay in how the swarm responds to the marker.

\section{Methods}
\subsection{Artificial diurnal lighting system}
The electrical circuit for the diurnal lighting setup consisted of a triac switch (TIC206, Bourns Inc, USA) connected to a microcontroller (Arduino Uno, Arduino, Italy), which regulated the flow of current to a 60 Watt incandescent bulb. Two opto-couplers (4N25, MOC3021, Fairchild Semiconductors) were used to isolate the main power supply from the microcontroller circuit. A real-time clock connected to the microcontroller (DS1307, Dallas Semiconductor, USA) was used to mimic diurnal lighting cycle, which consisted of 10.5 hours of day and night times corresponding to the bulb lit at full intensity and switched off. Dawn and dusk were simulated by gradually increasing and decreasing the light intensity between the day and night intensities over a 1.5 hour period. 

\subsection{Mosquitoes}
Mosquitoes for the experiments were obtained from a cyclic colony of \emph{An. stephensi} maintained in the insectary at the National Insitute of Malaria Research, Dwarka, India. The temperature and relative humidity at the insectary were maintained at 27$\pm$1$^{\circ}$ and 70$\pm$5\% respectively. Eggs were obtained from insect colony in a petri dish containing water and lined with blotting paper on the inner side. Eggs were allowed to hatch and were then transferred to enamel trays containing de-chlorinated tap water. Larvae were fed upon a mixture of powdered yeast and dog biscuits in a 60:40 ratio. The water in the rearing trays was replaced on a daily basis to avoid formation of scum.

\subsection{Initial experiments to assess indoor swarming}
\begin{figure}[h]
\centering
\includegraphics[width=0.5\linewidth]{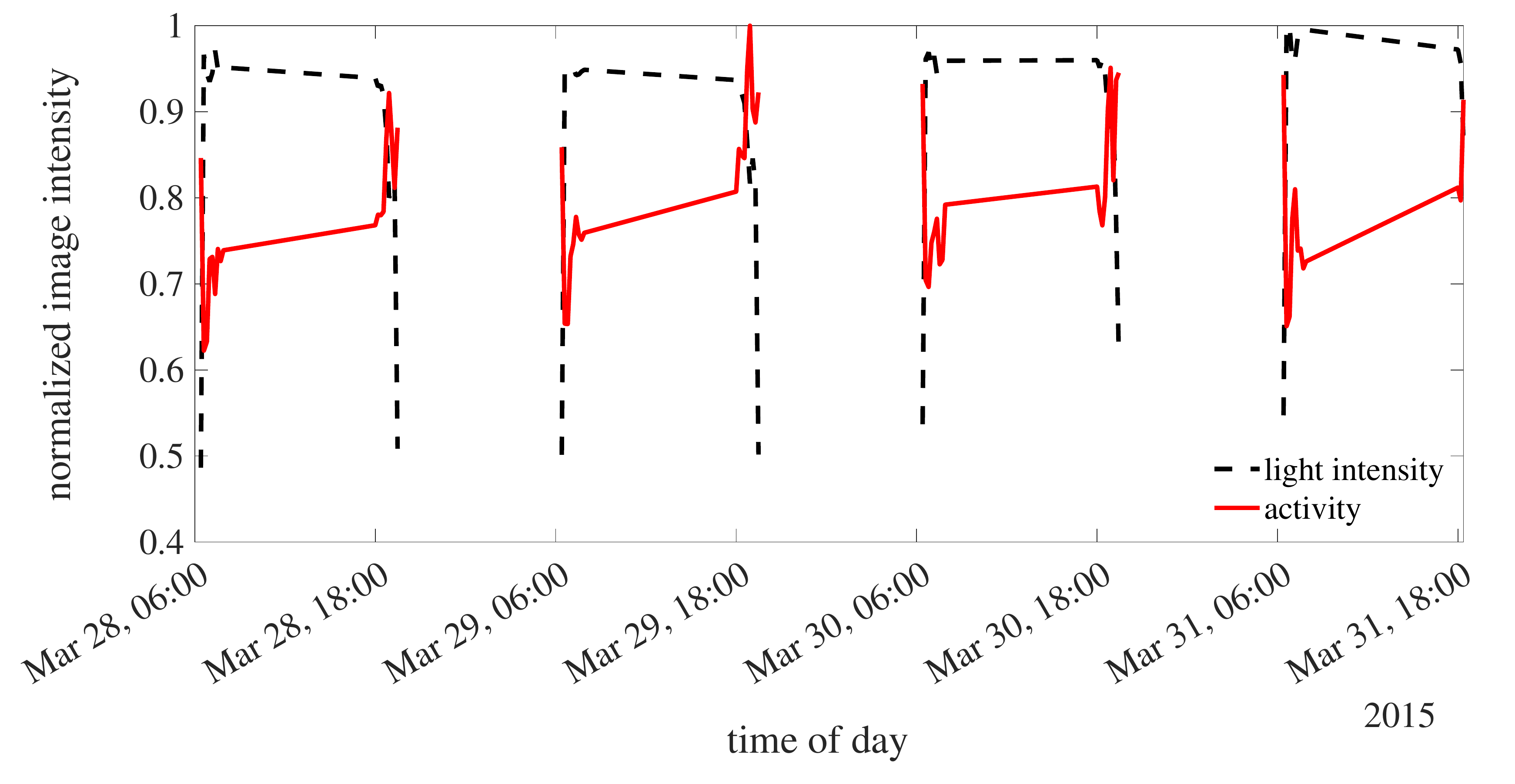}
\caption{Normalized swarm activity and ambient light intensity measured over four days}
\label{fig:preliminary_exp}
\end{figure}

Experiments were conducted to first assess whether \emph{An. stephensi} mosquitoes were capable of swarming indoors. Accordingly, bowls of mosquito larvae were put in a netted cage during the week beginning on March 28, 2015. The artificial diurnal lighting system was set up near the cage and a single camera (LifeCam Cinema 720p, Microsoft Corporation, USA) was set to record mosquito activity for four continuous days. The camera was mounted on a tripod and connected to a laptop computer (Asus X200MA-Bing-KX395B 11.6-inch Laptop, ASUS, India) with a solid state drive (ADATA Premier Pro SP600 SATA III MLC Internal Solid State Drive, 128GB, ADATA, USA) to record frames two times per day during the time at which the light bulb was dimmed or brightened. A custom shell script was run to record 15 frames for one minute every ten minutes during this time.

The images were processed to assess ambient light intensity and mosquito activity. Specifically, the average pixel intensity across a sequence of images during a minute was used to indicate ambient light intensity; swarming activity was measured in terms of the average difference in pixel intensity between successive images during the same minute. A sequence of images where the mosquitoes stayed in place would therefore be same and the pixel difference would be zero thereby recording no activity; swarming would result in successive images being different indicating high activity. Figure \ref{fig:preliminary_exp} shows the two indicators, light intensity and activity, normalized by dividing by the maximum observed value, over a period of four days. We found that the mosquitoes consistently swarmed during both dusk and dawn periods simulated by the diurnal lighting system.

\subsection{Experimental setup}
\begin{figure}[h]
\centering
\includegraphics[width=.5\linewidth]{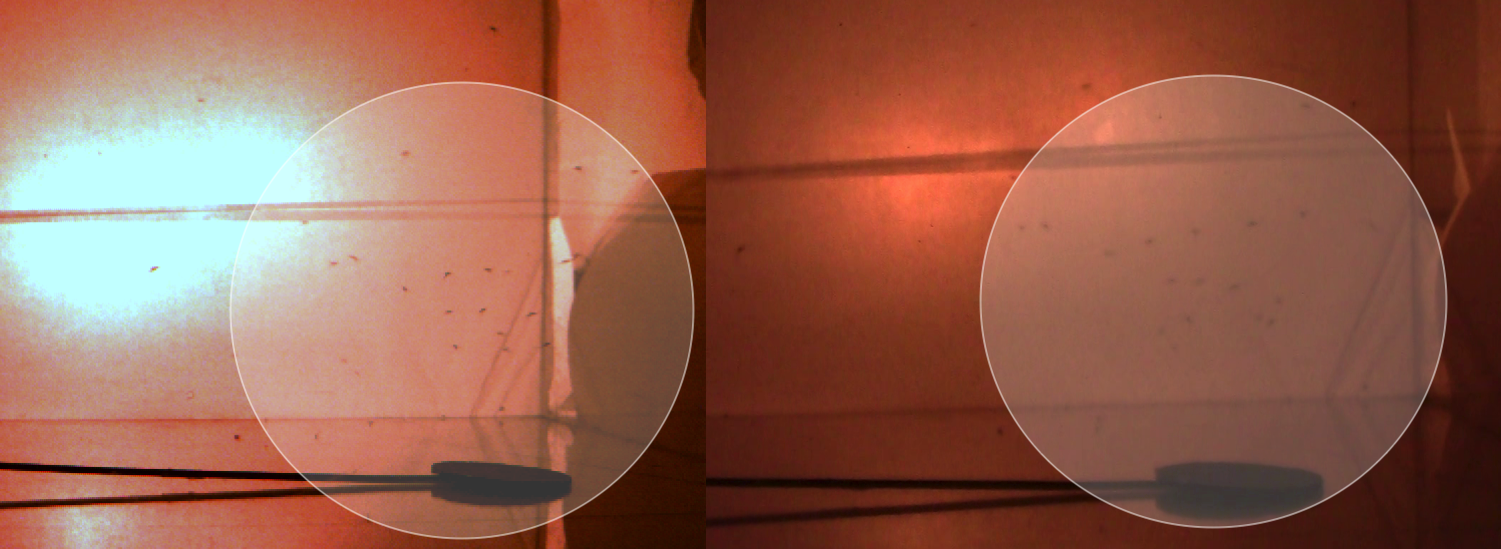}
\caption{Left and right camera image of a mosquito swarm over the marker. A circular region of the image where a majority of the swarm is present is highlighted.}
\label{fig:setup}
\end{figure}
The experimental setup consisted of a clear acrylic cage, cube with 61 cm side, and a stereo camera setup for filming the mosquitoes. The diurnal lighting system was placed 20 cm from on one side of the clear acrylic cage. The stereo camera system was placed facing the opposite side of the cage so that a back-lighting effect was created (see Supplementary figure 1).
The cage had two circular holes with diameter 15.2 cm, one each on opposite sides for ventilation, cleaning, and placing moist raisins as food. Nets were used to cover the two holes and was set in place with a 3D printed plastic ring. The side of the cage facing the bulb was covered with parchment paper sheet to diffuse the light and avoid glares. A hole in the side of the cage was made to permit moving the marker from outside the cage with an attached stick. 

The stereo camera setup consisted of two cameras (Flea3, Edmund Optics, Singapore, recording 1280 $\times$ 1024 pixel resolution frames at 60 frames per second, and a Canon Vixia HFR500 recording 1920 $\times$ 1024 at 30 frames per second) mounted on a custom stereo bar and a tripod (Gorillapod, Vitec Imaging Inc, USA). The cameras were mounted approximately parallel to each other with a baseline of 23 cm. Both cameras were calibrated using a calibration toolbox (MATLAB Calibration toolbox \cite{BouguetCalibrationToolbox}) to calculate the intrinsic (focal length and camera center) and extrinsic (relative position and orientation) parameters of each camera.

We used black cardboard discs as markers for the experiments (Fig.\ref{fig:setup}). A marker was attached to a 1 m long white colored stick made of bamboo through the circular hole. The stick was made to pass through a smaller hole on the side of the cage big enough to allow horizontal movement of the marker with little noise. Two markings on the stick were made to keep the moving distance constant within the cage. We tested three different marker sizes at 4,8, and 16 cm in diameter, to assess ease of use in terms of moving the marker back and forth and the ability to attract large swarms. We finally selected the 8 cm marker for experiments.

\subsection{Experimental Procedure}
Three days prior to an experimental trial approximately sixty male \emph{An. stephensi} mosquitoes were introduced into the cage in a bowl. Male mosquitoes were selected as follows. Once larvae were transformed into pupae (approx 7th day of the hatching of larvae), they were transferred in a plastic bowl containing water and placed into an insect cage, having access to 10\% glucose or water-soaked raisin, for emergence into adult mosquitoes. Since male mosquitoes emerged first, the early emergents were used for this study. Individual mosquitoes were sucked in a glass aspirator and checked for their sex under the illumination of a table lamp based on the physical appearance of antennae and palpi (the antennae of male mosquitoes are bushy and palpi are club-shaped). Confirmed male mosquitoes were transferred to the experimental cage.

A petri dish with raisins dipped in water was placed at the corner of the cage were put in the cage for feeding. On the day of the experiment, the petri dish was removed thirty minutes prior to filming. Although mosquitoes swarmed both during dawn and dusk periods, the experiments were only carried out during the dusk period. The experiments were conducted on a single batch of mosquitoes on March 25th and 26th, 2016. 

The experiment involved moving the marker back-and-forth along a single direction by hand with the attached stick. Markings on the stick were used to visually assist the experimenter to maintain a constant distance travelled within the cage. A checkerboard pattern was placed in the view of both cameras before the start of trials each day. This was done to perform extrinsic calibration separately for each day in the event that the cameras had moved.

The experimental procedure involved moving the marker back and forth for about a minute between two fixed points on the base of the cage, and then waiting for three minutes before moving the marker again with a different frequency. Care was taken to ensure that the marker movement did not cause any noise or movement of the cage. The different frequencies ranged between moving the marker very slowly between the two extreme locations to moving it very fast so that the swarm could not follow it at all. 
A stopwatch was used to maintain the frequency of marker movement. Out of the 18 total trials, five trials were selected for analysis based on the variability in marker driving frequencies.

\subsection{Analysis}
\subsubsection{Reconstructing three dimensional trajectories from stereo videos}
We processed the videos from each camera to extract frames.  Frames from the camera with the lower frame rate of 30 frames per second (with respect to 60 frames per second with the faster camera) were repeated so that the total number of frames for a given time were the same. Frames were then synchronized manually by ensuring that the marker movement and mosquito movement in the two frames were aligned. This was confirmed by creating a stereo video and watching it for the length of the video to ensure that no delay between the two camera views was observed (see \url{https://youtu.be/o5f383Unmas}). We note that the maximum delay that such an approach would introduce between two successive frames is 1/60th of a second. At a maximum speed of 1 m/s this would produce a reconstruction error of approximately 1.6 cm which is within one mosquito body length.

The synchronized stereo video data was then used to reconstruct three-dimensional trajectories of individual mosquitoes and the marker using a probabilistic multi-target tracking algorithm \cite{Butail2012b}. The algorithm was implemented in two stages. First, an automatic version ran a multi-hypothesis particle filter which reconstructed individual mosquito motion until it was lost due to occlusion between two or more mosquitoes that could not be resolved automatically, or sudden brightness intensity change because the camera sampling rate of 60 frames per second matched the 50 Hz frequency of the alternating current supply used to power the light source; in these situations the track was re-initiated with a different identity. The output of this first stage were a series of tracklets of length $4.5\pm2.5$ frames having different identities. Next, a graphical interface was used to manually verify and stitch together the tracklets into longer trajectories. The graphical interface also ran a particle filter to estimate three-dimensional position and velocity of a target based on user inputs. The final trajectories were smoothed using a Kalman filter and transformed so that three orthogonal axes corresponded to that along the direction of the marker and parallel to the image plane, along the vertical direction, and along the camera optical axis. 

To ensure consistent analysis, we selected a length of video within each dataset that showed the marker moving for at least one back-and-forth cycle. All mosquitoes swarming over the marker were tracked from the beginning to the end of the selected length of video until they either disappeared from view or settled on the wall of the cage.  

\subsubsection{Frequency Response Analysis}
The periodic movement of the swarm in response to the marker motivated a frequency response based analysis. Specifically, we computed the Fourier transform of individual mosquito and swarm movement in each direction to calculate the contribution of different frequencies. 
We compute the Fourier transform of the movement data of the swarm $X_{\omega}^{s}$, the marker $X_{\omega}^{m}$, and individual mosquitoes using the fast fourier transform (\emph{fft}) function in MATLAB. The absolute value and argument of the ratio of the Fourier transforms of swarm and marker denoted by $\hat{G}_c={X_{\omega}^{s}}/{ X_{\omega}^{m}}$ represents the relative gain and phase offset of the swarm to the marker at frequency $\omega$. 

A second aspect of our analysis aimed to characterize the coupling between the marker and the swarm. In classical control theory, the response $x(t)$ of a dynamical system to an input $u(t)$ can be described as
\begin{equation}
  a_0\frac{d^nx}{dt^n}+a_1\frac{d^{n-1}x}{dt^{n-1}}+\ldots + a_nx(t)=b_0\frac{d^mu}{dt^m} + b_1\frac{d^{m-1}u}{dt^{m-1}} + \ldots + b_mu(t),
\label{eqn:lti}
\end{equation} 
where $\frac{d^nx}{dt^n}$, for example, denotes the $n$-th derivative of $x(t)$ and is also known as the order of the model. This relationship described in time-domain can be represented in the frequency domain in the form of a transfer function
\begin{equation}
\bs{G}(s) = \frac{{X}(s)}{{U}(s)}
\end{equation}
where $s=j\omega$, $j=\sqrt{-1}$ and $\omega$ is a value of frequency at which the transfer function is evaluated; ${X}(s)$ and ${U}(s)$ are the frequency domain representation of ${x}(t)$ and ${u}(t)$ respectively. The transfer function is a complex number whose magnitude represents the ratio of magnitudes of the two signals, and whose argument represents the phase shift between two signals. A characterization of this relationship is called system identification \cite{Ogata1997}.
The process of system identification involves fitting the response in frequency or time domain of a system represented by (\ref{eqn:lti}) to the experimental values. Prior to system identification however, we must establish that the swarm response can indeed be represented by a linear dynamic relationship. Toward this we conducted linearity tests by computing coherence at a frequency $\omega$ defined by $C(\omega)=\frac{|R_{MS}(\omega)|^2}{R_{MM}(\omega)R_{SS}(\omega)}$, where $R_{MS}$, is the cross power spectral density between the marker and the swarm, and $R_{MM}$, and $R_{SS}$ are the power spectral density of the marker and swarm respectively. A high coherence value close to 1 indicates linear relationship \cite{Sponberg2015}.

\begin{figure}[th]
\centering
\includegraphics[width=.995\linewidth]{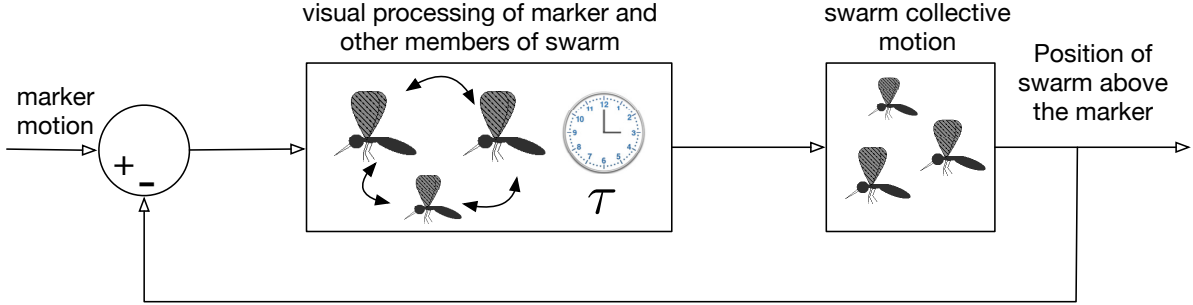}
  \caption{We model the swarm response to the marker in the form of a feedback loop. Individual mosquitoes sense relative motion of the marker and members of the swarm and process it to give rise to collective behavior of following the marker. More mosquitoes in a swarm implies more social interactions which we model in the form of processing delay of the marker motion.}
\label{fig:feedback_loop}
\end{figure}
We denote the swarm dynamics in the form of a transfer function with an internal time delay $\tau$
$G_o(s)e^{-s\tau}$.
Since the swarm manages to follow the marker as it changes direction, we consider the existence of a feedback loop \cite{Cowan2014}, where each mosquito tends to correct its motion based on how it senses the environment and other mosquitoes. Accordingly, the swarm response to the marker can be modeled as a feedback loop as in Figure \ref{fig:feedback_loop}. Within this framework, the swarm responds to the marker as it processes social interactions within, which give rise to an internal time delay. The overall transfer function can therefore be written as \cite{Sponberg2015}
\begin{equation}
G_{c}(s)=\frac{G_o(s)e^{-s\tau}}{1+G_o(s)e^{-s\tau}}.
\end{equation} 
System identification is performed by minimizing the sum $\min_{G_o(s), \tau}\sum_{i=1, \ldots, 5}\lvert G_c(s_i)-\hat{G}_c(s_i)\rvert$ over possible values of $G_o(s)$ and a single value of $\tau$ within the \emph{fmincon} routine in MATLAB. To test the dependence on swarm size, we additionally let the delay vary for each experiment so that the minimization is performed over possible values of $G_o(s)$ as well as a variable internal time delay as $\min_{G_o(s), \tau_i}\sum_{i=1, \ldots, 5}\lvert G_c(s_i)-\hat{G}_c(s_i)\rvert$.

\section{Results}

\subsection{Temporal dynamics of mosquito response to marker movement}
\begin{figure}[th]
\centering
\includegraphics[width=.995\linewidth]{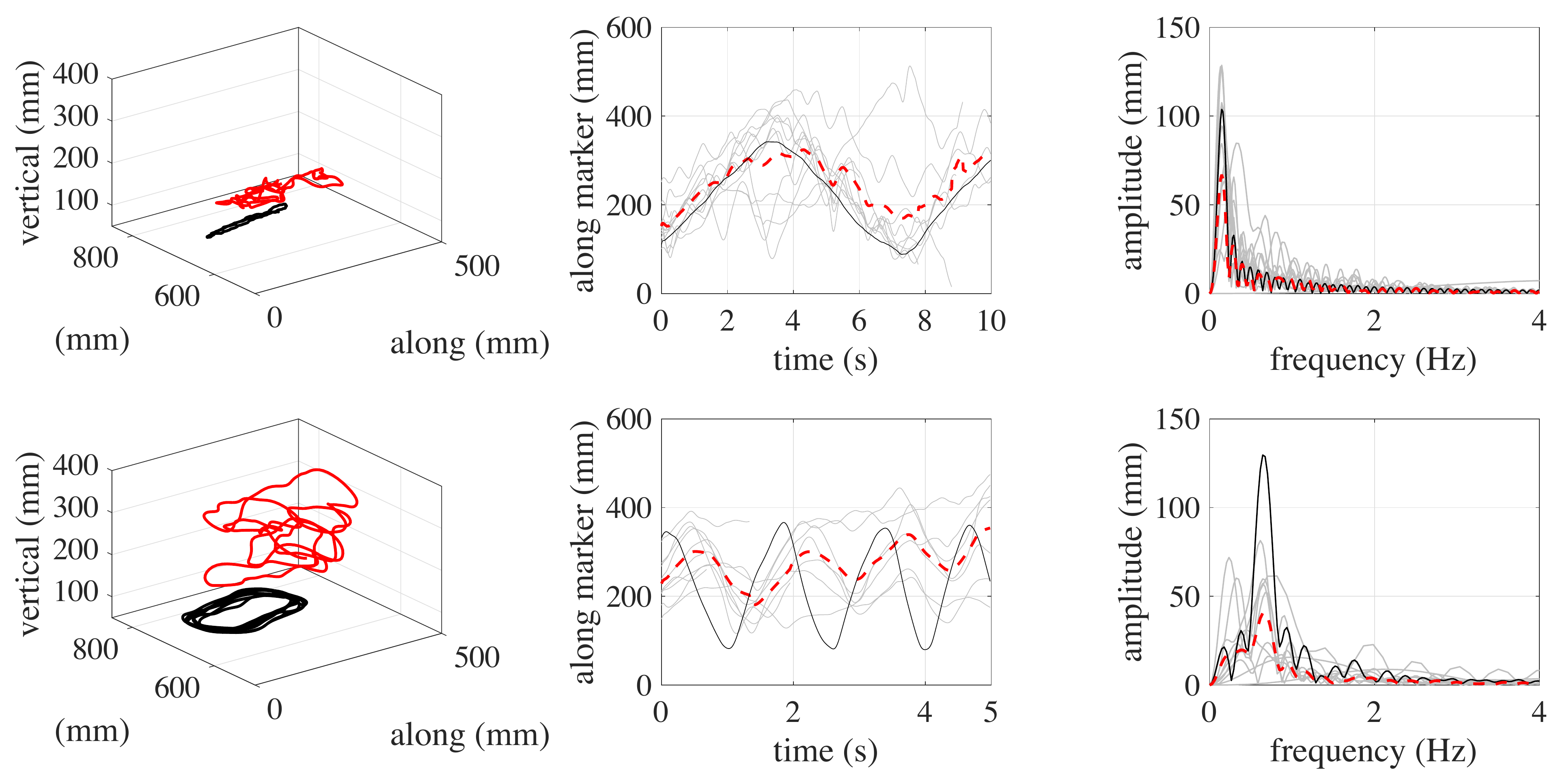}
  \caption{Trajectory of a single mosquito (red) as the marker moved (black) at two different frequencies (left column). Position of all mosquitoes (grey) and the swarm centroid (dashed red) along the direction of the marker; the marker position is shown in black (middle column). The frequency responses of the swarm centroid (dashed red), marker (black) and individual mosquitoes (grey) are shown in the direction along the movement of the marker (right column).}
\label{fig:summary_fig}
\end{figure}


Figure \ref{fig:summary_fig} shows the movements for the marker and the tracked mosquitoes in three-dimensions and then along the direction of marker movement for the lowest and highest frequency. (See Supplementary figure 2 for all five datasets.) The combined data for this study represents 16,241 three-dimensional position and velocity estimates from 63 trajectories. 

We find that the primary movement of the swarm is along the direction of movement of the marker. In the case, when the marker was moved at highest frequency, a distinct back-and-forth movement was also present in the direction normal to the marker which the swarm also tended to follow.  The velocities of the swarm also followed a periodic pattern (Supplementary figure 3).

\subsection{Frequency response analysis of swarming over the moving marker}

Figure \ref{fig:summary_fig} also shows the frequency response characteristics of the marker-swarm interaction along the direction of the marker (all datasets shown in Supplementary figure 5). The dominant frequency (highest peak in each figure) denotes the frequency with which an entity (marker, swarm, or a mosquito) was primarily moving in the direction along the movement of the marker. We note that the dominant frequency for the swarm centroid aligns closely with that of the marker. Specifically, the dominant frequency for the swarm centroid (with respect to the marker) are 0.14 (0.14), 0.20 (0.20), 0.41 (0.41), 0.55 (0.52), 0.61 (0.64) Hz respectively. 
The contribution of the dominant frequency to the swarm response is relatively low for when the marker is driven at high frequencies (0.52 and 0.64 Hz) compared to the low driving frequencies (0.14, 0.20, and 0.41 Hz). 

\subsection{System identification of swarm-marker relationship}

\begin{figure}[th]
\centering
\includegraphics[width=0.995\linewidth]{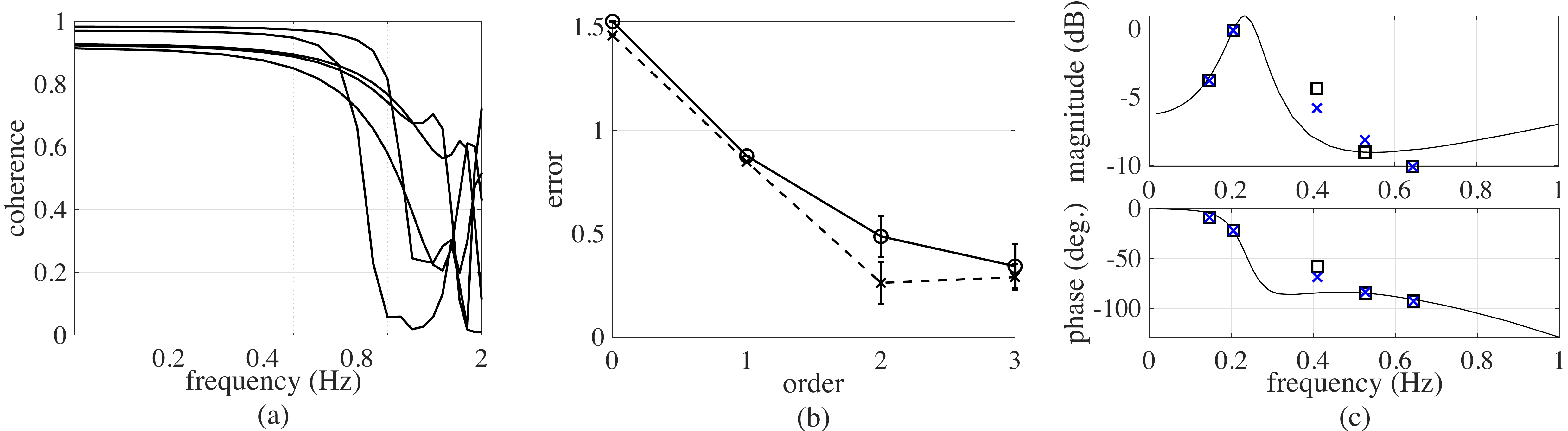}
  \caption{(a) Coherence of all five datasets with respect to frequency. (b) Error $\pm$ standard deviation in frequency response fit as a function the order of model for a model with constant delay (solid) and variable delay (dashed). The different values correspond to the different number of zeros that are possible for a given number of poles in the transfer function. (c) Ratio of magnitude and phase shift of the frequency response of the swarm to the marker (squares), best fit second order model with constant delay (solid line) and separate variable delay model for each frequency (crosses).}
\label{fig:model_fits}
\end{figure}
\begin{table*}[t]
    \centering
    	\begin{tabular}{@{}llll@{}}
    \hline
	  \bf Date \& time & \bf Driving freq. (Hz) & \bf Swarm size & \bf Internal time delay (s)\\
      \hline
	  2016-03-26 19:17:07  & 0.14 & 5.27 $\pm$ 0.44 & 0.0654\\
	  2016-03-26 19:14:54  & 0.20 & 5.74 $\pm$ 1.86 & 0.0402\\
	  2016-03-26 19:07:55  & 0.40 & 9.57 $\pm$ 2.24 & 0.6783\\
	  2016-03-25 19:17:17  & 0.52 & 9.42 $\pm$ 2.28 & 0.7891\\
	  2016-03-26 19:20:11  & 0.64 & 7.80 $\pm$ 1.12 & 0.0718\\
    \end{tabular}
  \caption{Details of the trials that were analyzed for this study in the order of increasing marker frequency. Swarm size denotes the average number of positions tracked. The internal time delay represents the intra-swarm processing of external cues in the linear system representation of the swarm-marker relationship.}
    \label{tab:data}
\end{table*}
The coherence between swarm and marker movement was found to be more than 0.8 for the range of frequencies with which the marker was moved (Fig. \ref{fig:model_fits}a). Identification of the linear dynamical system that best represents the relationship between the marker and the swarm revealed that the error reduced as we increased the order of the system from 0 (a proportional gain) to 4. The reduction in error is little beyond a second order system (h1) which has a non-zero delay of 0.354 s (h2, Fig. \ref{fig:model_fits}b). A variable delay model produces less error than a constant delay model. Figure \ref{fig:model_fits}c shows the relative magnitude and phase of the response along with the best fit second order model with constant delay and variable delay (See Supplementary material for transfer function forms of the swarm dynamics). 
Table \ref{tab:data} lists the five trials with the average number of positions tracked and the amount of variable delay. 
The amount of delay correlates to the average number of positions tracked (h3, linear regression model, $p=0.0494, F=10.2, R^2=0.773$). No correlation exists between the amount of delay and driving frequency (linear regression model, $p=0.525, F=0.513, R^2=0.145$).

\section{Discussion}
Swarming in insects is a complex collective behaviour governed by multiple internal and external stimuli. In some insects that fly, a visual marker forms a crucial external stimulus for initiating and maintaining the swarming behavior in the wild \cite{Dublon2014}. Here, we conducted laboratory experiments with a \emph{An. stephensi} mosquito swarm responding to a moving marker. The swarm followed the marker as it moved to-and-fro with different frequencies. A mathematical relationship representative of a feedback loop was fit to the frequency response data of the swarm.

\paragraph{Swarming indoors with artificial lighting} \emph{An. stephensi} mosquitoes were found to swarm in a cage during low light conditions induced artificially by an electronic lighting system. The swarming behaviour was consistently observed for five continuous days and provided the basis for conducting marker based experiments described here. Although we did not put a marker in the cage during our preliminary experiments, it is likely that the larvae bowls acted as markers for the swarming mosquitoes. Since the larvae consisted of both males and females, it is also possible that the mosquitoes that sat on the sides of the cage were predominantly female. Since the swarms were observed at both times during which the lighting was dimmed it is likely that the light intensity played a primary role in inducing a swarm.

\paragraph{Marker following behavior} In following the marker, the swarm demonstrated a distinct repetitive movement pattern along the direction of movement of the marker. It is unlikely that this movement is due to any external influence other than the marker. This is because at the highest frequency of 0.64 Hz in our experiments, the hand movement was not entirely parallel to the camera plane and in that case, we simultaneously find that the swarm follows the marker in a direction normal to the camera plane. It is also unlikely that the hand holding the marker stick created a looming stimulus, which would constitute a repelling influence only, and would not have attracted the swarm towards itself when moving away from the cage.

\paragraph{Hypotheses related to system identifcation} A second order description of swarm dynamics within a feedback loop serves well to represent the swarm response to the marker. This agrees with similar findings in midges \cite{Okubo1986, vanderVaart2019} and mosquitoes \cite{Butail2013}.    
The amplitude ratio and phase difference between the marker and the swarm movement show that the swarm is best able to keep up at a frequency of 0.2 Hz where it has nearly the same amplitude and a low phase difference. Interestingly, this is not the lowest frequency at which the marker was moved indicating the presence of an ideal optic flow from the environment. At 0.64 Hz, the highest frequency at which we moved the marker, the magnitude ratio was close to 0.4 showing that the swarm was only able to keep up 40\% of the way and lagged behind at nearly half a cycle.

\paragraph{Dependence of delay on swarm size} The existence of a non-zero internal time delay confirm our hypotheses about how the swarm interacts with the marker. The existence of correlation between time delay and swarm size further confirms that the interactions within the swarm need to be balanced against external cues, much like in other animal groups. Compared to \cite{vanderVaart2019}, where midge swarms were shown to respond viscoelastically to external perturbations, with social interactions likely manifesting in the form of elasticity of the swarm, the modeling framework here imparts a single description (transfer function) to the swarm over varying frequencies and seeks to model social interactions in the form of an internal delay. We did not find a frequency dependence of the internal delay; finding such a dependence would imply that the internal dynamics of the swarm vary with driving frequencies---a result that may have been expected in light of the dependence of storage and loss moduli on driving frequencies in midge swarms \cite{vanderVaart2019}. However, a direct relationship between an internal time delay as modeled here and viscosity of a swarm is not obvious. At the same time, the experiments described here are preliminary with a much smaller dataset and relatively smaller swarms, and should therefore only serve to motivate future hypotheses. 
The value of the delay is much higher at 0.678 s and more for swarms that have nine or more mosquitoes tracked per time step (the actual number of mosquitoes in the swarm are possibly more because we do not track mosquitoes that enter the swarm during the time we tracked) than 0.071 s for swarms where less than 8 mosquitoes per time step. This is a large difference in the delay with a small increase in the number of mosquitoes and suggests a nonlinear relationship warranting further investigation in terms of more experiments and larger swarms. 
Does the existence of a delay imply that a single mosquito would have followed the marker without delay? It is difficult to answer this question because the social interactions may also serve to reinforce the response to an external stimulus. Isolating environmental versus social cues in collective behavior is an open question in collective animal behavior \cite{Miller2013}, and methods in information theory have been particularly effective in separating causal influences across space and time \cite{Lord2016inference, Butail2019detecting}. Future experiments with swarming insects that can help tease out the effects of external versus internal information will require controlling the environment at a higher detail with multiple perspectives and modalities. With respect to mosquitoes, acoustics have been found to play a critical role in species and sex recognition \cite{Simoes2016}, and therefore isolating environmental and social effects entails an accurate reproduction of sound within the swarm. In this context, an accurate sound propagation and perception model must be developed for implementation on top of three-dimensional position data. Such a model can be used to reconstruct the visual and acoustic cues perceived by individual mosquitoes within a swarm. Finally, the robustness properties of the marker following behavior can inform target tracking algorithms for robotic swarms \cite{Senanayake2016search}.

\paragraph{Vector control} From the perspective of vector control, results in this study provide a quantifiable description for evaluating landmarks which in the form of markers, act as consistent mating sites \cite{Sawadogo2017}. For example, field observations of the \emph{An. gambaie ss.} have indicated that swarms generally form over high contrast regions that interrupt the regularity of the landscape \cite{Diabate2015}. The results here imply that in the wild a swarm of \emph{An. stephensi} mosquitoes would be able to follow moving markers (e.g. head of a human or an animal)  at speeds of up to 50 cm/s before they are lost. Comparing different markers with different species of mosquitoes \cite{Poda2019} in terms of the type and parameters of the visual coupling such as coefficients of the second order system can help in landscape design that discourages the formation of swarms in disease prone regions.

\section*{Acknowledgments}
Thanks to 
Mr.~Bhoopal and Mr.~C.~Dwivedi for help with the experiments. Dr.~Derek Paley originally proposed a similar experiment in 2011 as part of a field study. We also thank anonymous reviewers for their constructive inputs on an earlier draft of this manuscript.

\bibliography{references}
\bibliographystyle{ieeetr}

\end{document}


\maketitle
\renewcommand{\figurename}{Supplementary Figure}

\begin{figure}[th]
\centering
\includegraphics[width=0.495\linewidth]{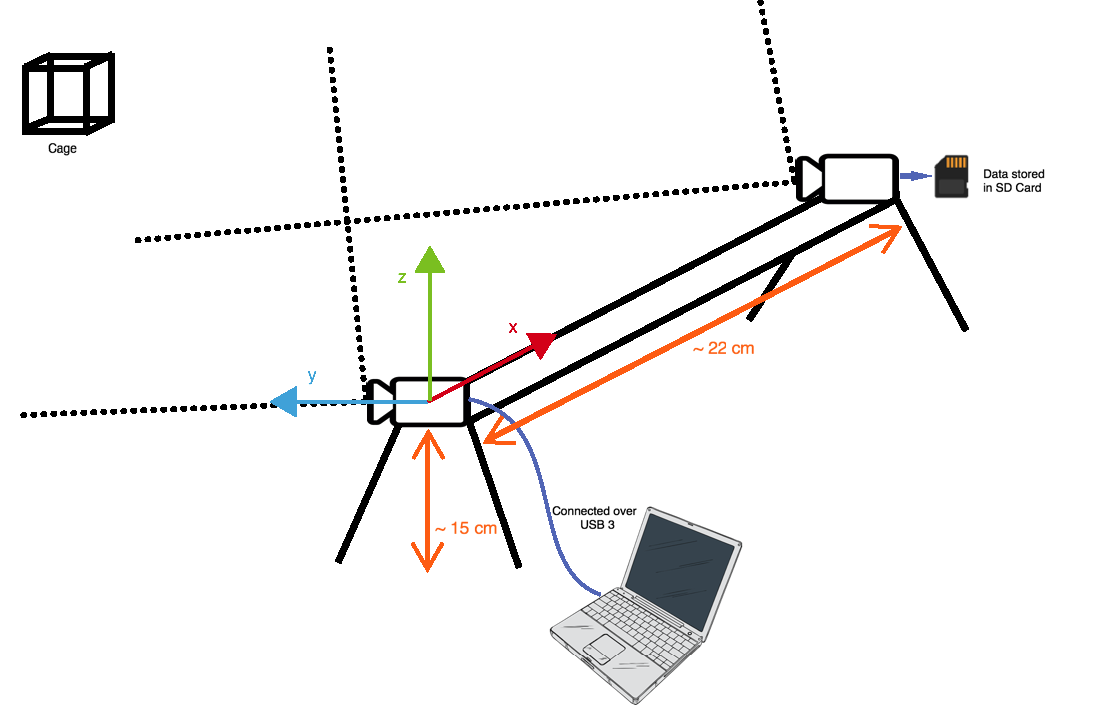}
\includegraphics[width=0.495\linewidth]{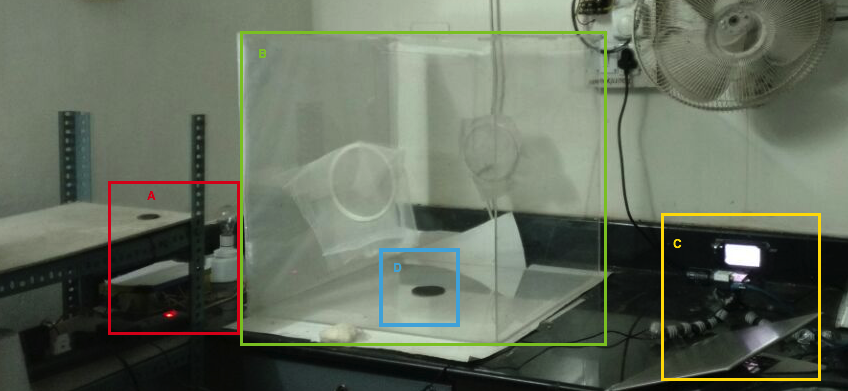}
\caption{Schematic (left) and a picture (right) of the experimental setup. The experimental setup consisted of a diurnal lighting system (A) that automatically backlit an insect cage (B) as it was filmed with stereo camera system (C) mounted in front of a cage. The black marker (D) used in the experiments is also shown.}
\label{fig:exp_setup}
\end{figure}



\newpage
\begin{figure}[th]
\centering
\includegraphics[width=.995\linewidth]{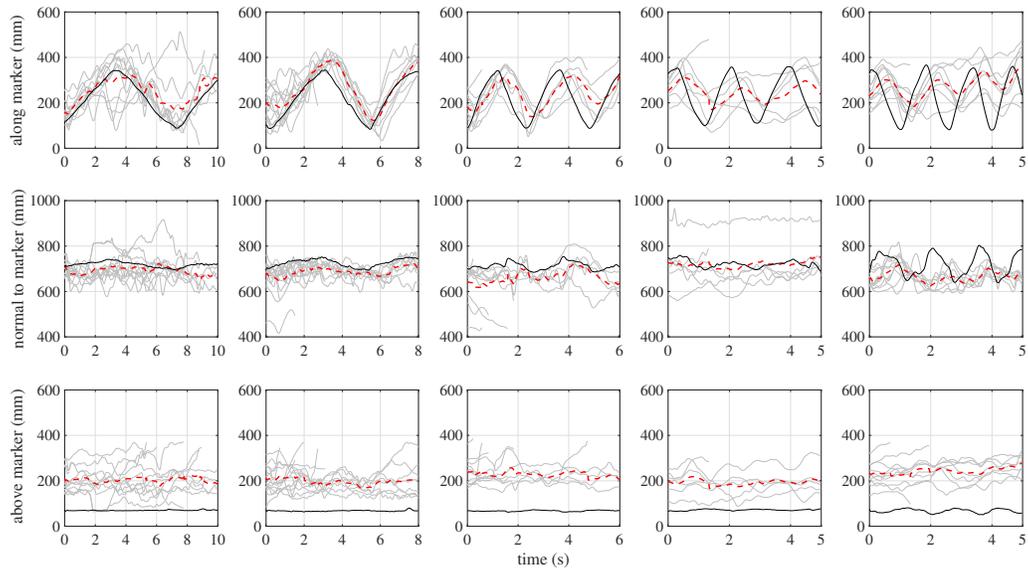}
\caption{Position of all mosquitoes (grey) and the swarm centroid (dashed red) along the three directions as the marker moved back and forth. The marker position is shown in black.}
\label{fig:pos_plot}
\end{figure}

\newpage
\begin{figure}[th]
\centering
\includegraphics[width=.995\linewidth]{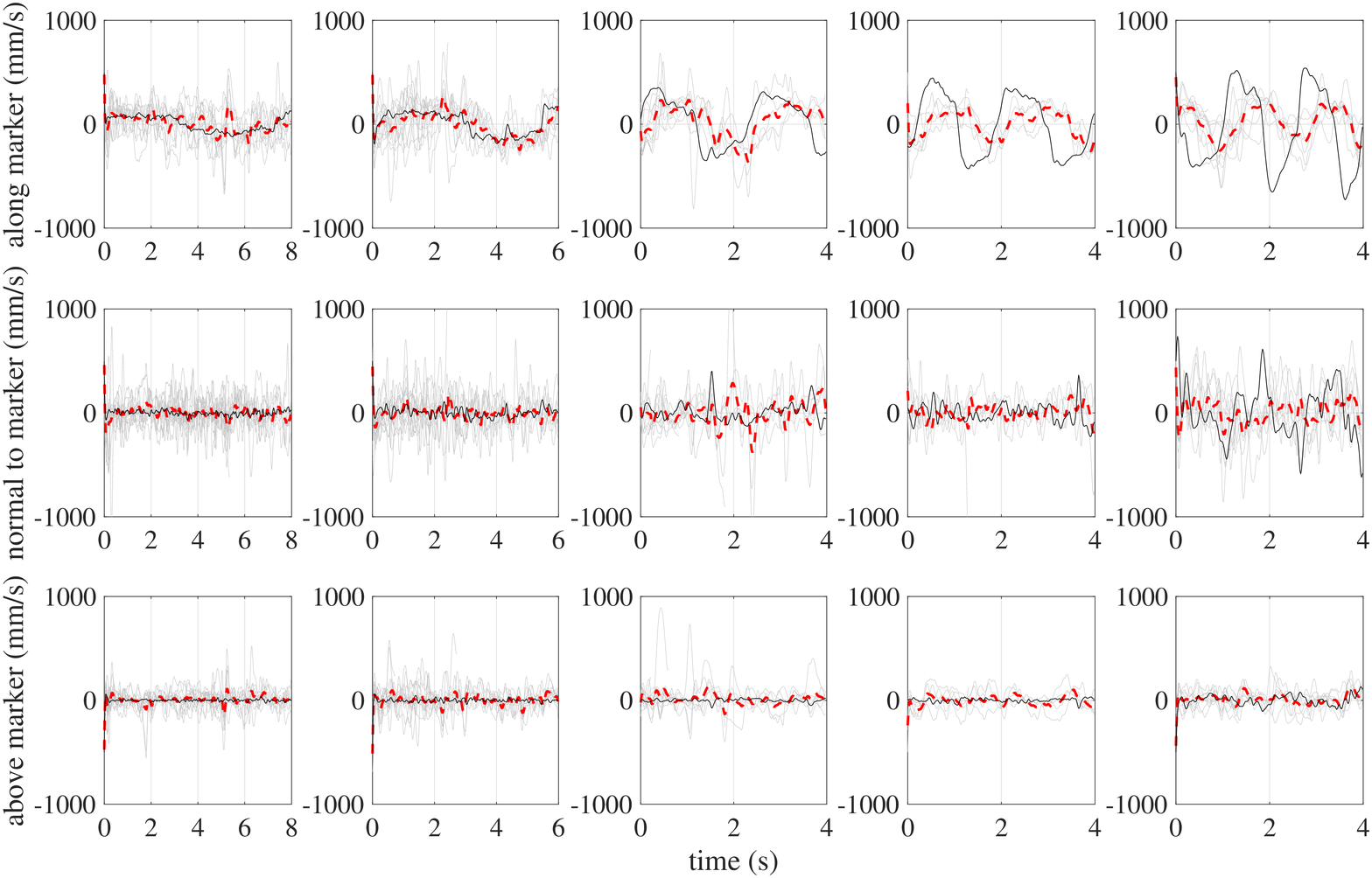}
\caption{Velocities of all mosquitoes (red) within a swarm along three independent directions as the marker moved back and forth. The marker position is shown in black.}
\label{fig:main_vel}
\end{figure}

\newpage
\begin{figure}[th]
\centering
\includegraphics[width=0.995\linewidth]{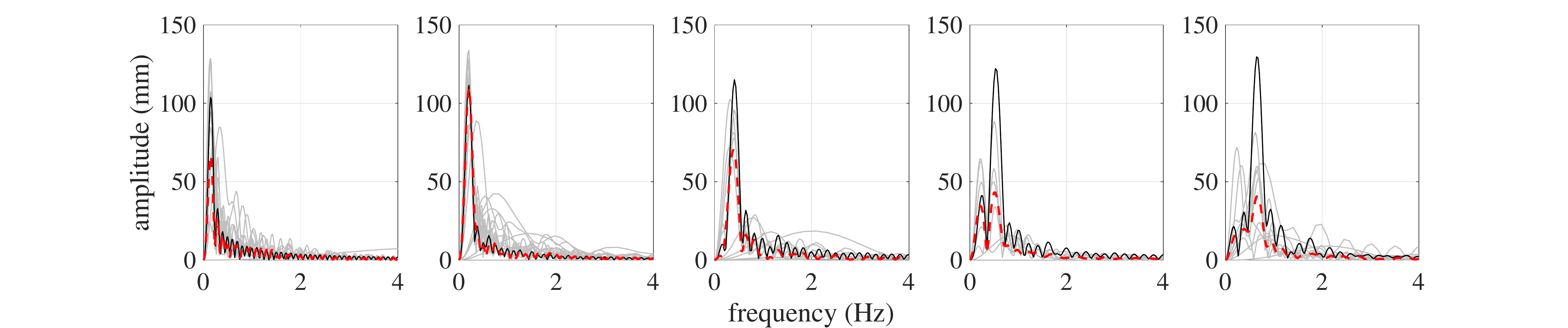}
\caption{The frequency responses of the swarm centroid (dashed red), marker (black) and individual mosquitoes (grey) are shown in the direction along the movement of the marker. The dominant frequencies (left to right) of the marker are 0.14, 0.20, 0.41, 0.52, and 0.64 Hz.}
\label{fig:main_fft}
\end{figure}

\newpage
\begin{tcolorbox}[float=htb, title=Transfer functions]
The transfer function for open-loop dynamics with constant delay that best fit the data was
\begin{equation}
\begin{split}
  G_o(s)&=\frac{0.3481 s^2 + 0.7023 s + 1.379}{1.023 s^2 + 0.3121 s + 1.452}e^{-0.354s} \\
  &= \frac{(s+1-1.715 j)(s+1+1.715j)}{(s+0.1525-1.1813j)(s+0.1525+1.1813j)}e^{-0.354s}
  \end{split}
\end{equation}
where $j=\sqrt{-1}$.  The open-loop transfer function with variable delay was
\begin{equation}
\begin{split}
  G_o(s)&=\frac{1.185 s + 0.8562}{1.1 s^2 + 0.4979 s + 1.596}e^{-\tau_is} \\
  &= \frac{(s+0.7224)}{(s+0.2262-1.1827 j)(s+0.2262+1.1827j)}e^{-\tau_is}
  \end{split}
\end{equation}
where $\tau_{i}$ corresponds to the delay corresponding to the fit for a trial $i$. 
\label{tbox:open_tfs}
\end{tcolorbox}